\documentclass[iop,apjl,twocolumn]{emulateapj_pab}
\RequirePackage{hyperref}
\hypersetup{
	colorlinks = true,
	linkcolor = blue,
	citecolor = blue,
	filecolor = blue,
	urlcolor = blue,
	pdfborder = {0 0 0}
}
\usepackage{color}
\graphicspath{{./}{Bilder/}}

\renewcommand*\aap{A\&A}

\renewcommand*\ao{Appl.~Opt.}

\renewcommand*\apjl{ApJL}

\renewcommand*\apjs{ApJS}

\newcommand{\sect}[1]{Section~\ref{S:#1}}

\newcommand{\fig}[1]{Figure~\ref{F:#1}}

\newcommand{\eql}[1]{\begin{equation}#1\end{equation}}

\newcommand{\eqi}[1]{$#1$}

\DeclareRobustCommand*{\unit}[1]{\def~{\,}\ensuremath{\mathrm{\,#1}}}
\definecolor{darkgreen}{rgb}{0,0.45,0}

\begin{document}
\hypersetup{
	pdftitle = {Plasma Beta Stratification in the Solar Atmosphere: A Possible Explanation for the Penumbra Formation},
	pdfauthor = {Ph.-A.~Bourdin},
	pdfkeywords = { magnetic fields -- methods: numerical -- Sun: chromosphere -- Sun: corona -- Sun: photosphere -- sunspots},
	pdfsubject = {The Astrophysical Journal Letters, 850(2017) L29. doi:10.3847/2041-8213/aa9988}
}


\title{Plasma Beta Stratification in the Solar Atmosphere:\\A Possible Explanation for the Penumbra Formation}
\shorttitle{Plasma beta in the solar atmosphere}
\shortauthors{Bourdin}


\author{Ph.-A.~Bourdin \href{https://orcid.org/0000-0002-6793-601X}{\includegraphics[width=0.3cm]{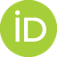}}}
\affil{Space Research Institute, Austrian Academy of Sciences, Schmiedlstr. 6, A-8042 Graz, Austria, \href{mailto:Philippe.Bourdin@oeaw.ac.at}{Philippe.Bourdin@oeaw.ac.at}}



\received{2017 May 19}
\revised{2017 October 26}
\accepted{2017 November 8}
\published{2017 November 27}

\submitted{}
\journalinfo{The Astrophysical Journal Letters, {\rm 850:L29 (5pp), 2017 December 1 \hfill \url{https://doi.org/10.3847/2041-8213/aa9988}}}

\begin{abstract}
Plasma beta is an important and fundamental physical quantity in order to understand plasma dynamics, particularly in the context of magnetically active stars, because it tells about the domination of magnetic versus thermodynamic processes on the plasma motion.
We estimate the value ranges of plasma beta in different regions within the solar atmosphere and we describe a possible mechanism that helps forming a penumbra.
For that we evaluate data from a 3D~magnetohydrodynamic model of the solar corona above a magnetically active region.
We compare our results with previously established data that is based on magnetic field extrapolations and that was matched for some observations.
Our model data suggest that plasma beta in the photosphere should be considered to be larger than unity outside of sunspots.
However, in the corona we also find that the beta value range reaches lower than previously thought, which coincides with a recent observation.
We present an idea based on a gravity-driven process in a high-beta regime that might be responsible for the formation of the penumbra around sunspot umbra, where the vertical field strength reaches a given threshold.
This process would also explain counter-Evershed flows.
Regarding the thermal and magnetic pressure within the mixed-polarity solar atmosphere, including non-vertical magnetic field and quiet regions, plasma beta may reach unity at practically any height from the photosphere to the outer corona.
\end{abstract}

\section{Introduction} \label{S:intro}

The value of plasma beta is of high relevance for understanding the dynamics of the Sun, in particular if magnetic or thermodynamic processes dominate in that plasma.
Estimating $\beta$ in the solar atmosphere requires knowledge of the magnetic flux density $B$ that can currently only be measured with good accuracy by photospheric spectroscopy.

\cite{Gary:2001} conducted a 1D potential field extrapolation from magnetic flux estimates for sunspots and plage areas in the photosphere, together with solar atmospheric stratifications also taken from relatively old estimates based on inversion techniques \citep{Vernazza+al:1981} that consider only a rather strong, vertical magnetic field.
\cite{Gary:2001} artificially stretched the extrapolation to fit observations in the far-out corona and solar wind.

The 3D~magnetohydrodynamic (MHD) simulation we use for this work is driven by observed photospheric magnetograms of a small active region with surrounding quiet Sun -- and features a self-consistently computed magnetic field and plasma pressure in the corona; see \sect{simulation}.

For extrapolations of the coronal magnetic fields, be it potential or nonlinear force-free, one usually assumes $\beta \ll 1$.
\cite{Gary:2001} already finds that this regime does not extend very high up into the corona.
In \sect{beta}, we show that this assumption about plasma beta in the corona can be violated at any height.

In \sect{penumbra}, we sketch the formation of penumbra around sunspots through the strong variability in plasma beta that helps magnetic fields to change their inclination.
Observational findings already suggest that magnetic field lines fall from the chromosphere to the photosphere, in particular not near emerging magnetic flux; see \cite{Bello_Gonzales+al:2017}.

\section{MHD simulation} \label{S:simulation}
We performed a 3D~MHD simulation of the solar corona above a small and sunspot-less active region surrounded by a magnetically quiet region \citep{Bourdin+al:2013_overview}.
Together with the resistive and compressible MHD equations, we also implement the necessary energy sinks to obtain a realistic energy balance for the corona.
The isotropic, spatially uniform, and temporally constant magnetic resistivity \eqi{\eta} adds to the thermal energy, while a radiative-loss function \citep{Cook+al:1989} and Spitzer-type heat conduction \citep{Spitzer:1962} remove thermal energy.
Together, both set the coronal plasma density and temperature through a self-consistent magnetic evolution and energy conversion within the corona.

At the lower boundary of the simulation domain, we apply a line-of-sight magnetogram (observed on 2007~November~14 at 15:00--16:00 UTC near disk center by {\em Hinode} SOT/NFI) that we calibrated with the magnetic flux densities $B$ taken from {\em Hinode} SOT/SP magnetic vector data \citep{Kosugi+al:2007,Tsuneta+al:2008}.
The density $\rho$ and temperature $T$ that we impose on the lower boundary of our simulation domain match those values in the photosphere of the Sun.
Therefore, the thermal pressure $P_{\rm therm}$ (set by $\rho$ and $T$) and the magnetic pressure $P_{\rm mag}$ (set by $B$) are both fixed to known constraints in the photosphere -- and the plasma beta in this layer is hence consistent between simulation and reality.
Also, the atmospheric stratification above the photosphere is constrained by observed values \citep{Bourdin:2014_switch-on}.

We compare the modeled coronal EUV-bright structures and the plasma flows therein with spectra obtained by {\em Hinode}/EIS \citep{Culhane+al:2007}.
Stereoscopic observations performed by the {\em STEREO} satellites \citep{Howard+al:2008} match well with the 3D structure of the coronal loops in the model.
The coronal loops in the MHD model, and the plasma flows along them, reflect well their real counterparts \citep{Bourdin+al:2013_overview,Bourdin+al:2014_coronal-loops}.
Further analyses show this model heats the corona as required \citep{Bourdin+al:2015_energy-input} and it shows some similarities to earlier scaling laws for coronal plasma properties \citep{Bourdin+al:2016_scaling-laws}.

\section{Plasma beta} \label{S:beta}
The plasma beta is the ratio between the thermal pressure $P_{\rm therm}$ and the magnetic pressure $P_{\rm mag}$ that we compute from our MHD data as
\eql{\beta = {P_{\rm therm} \over P_{\rm mag}} = 2 \mu_0 c_p {\gamma - 1 \over \gamma} {\rho T \over B^2}}
with the magnetic flux density $B$, the plasma density $\rho$, the temperature $T$, the adiabatic index $\gamma$, the specific heat capacity at constant pressure $c_p$, and the magnetic permeability $\mu_0$.
Averages, minima, and maxima are taken in horizontal layers for each height.
Stronger magnetic field (at constant thermal pressure) implies a smaller plasma beta, which in the same time means that the plasma motion becomes dominated by the magnetic field configuration for $\beta \ll 1$ and, vice versa, magnetic field is dragged with the plasma bulk flow for $\beta \gg 1$.
When $\beta$ is near unity, both can affect each other.

\begin{figure}
\plotone{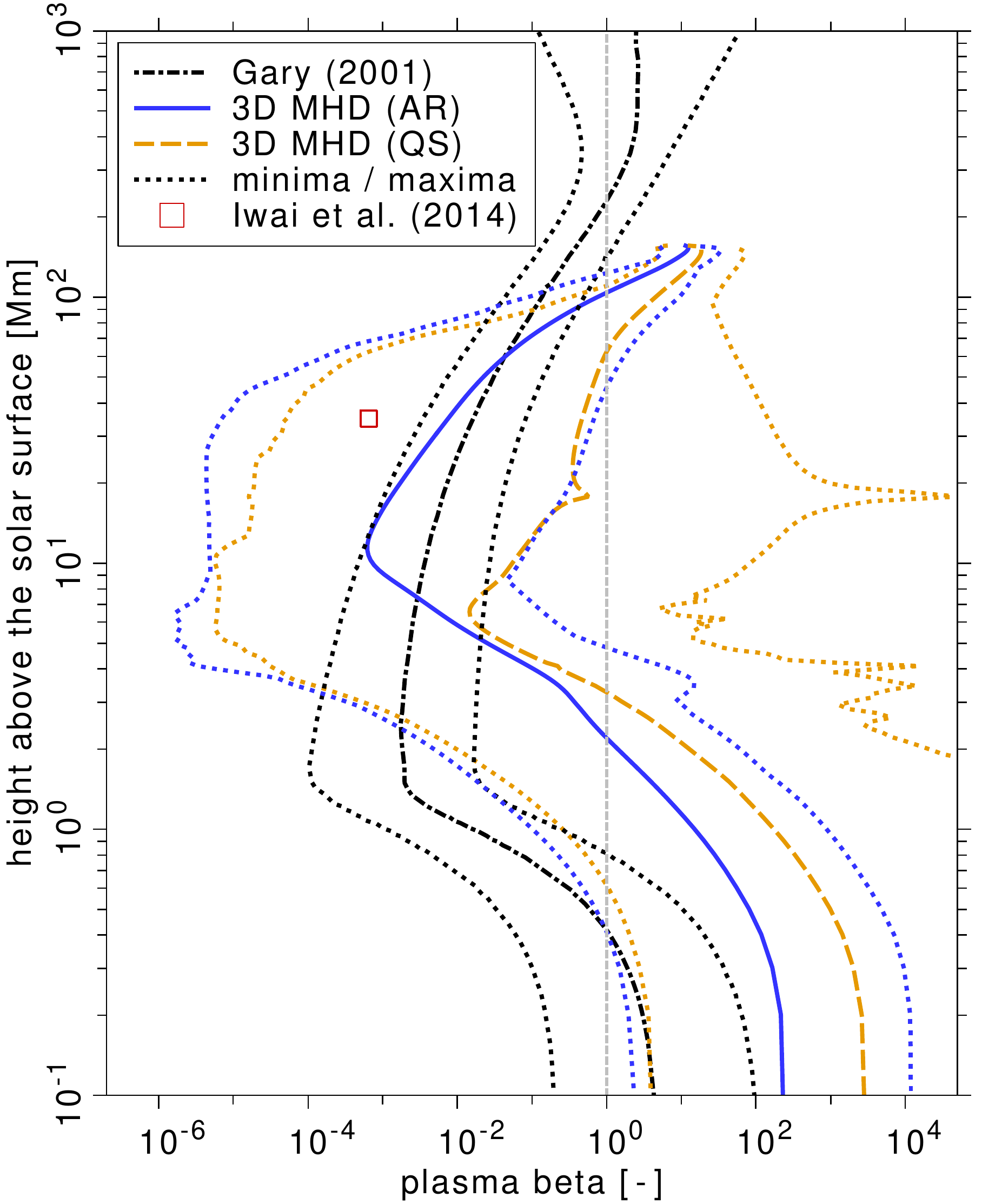}
\caption{Plasma beta versus height above the photosphere.
The blue solid line is the mean value for the active region (AR) core area, as indicated in \cite{Bourdin+al:2013_overview}.
The dotted lines indicate the minimum and maximum values within each height layer.
The orange data represent the complement to the AR core in our MHD model, which is some quiet Sun (QS) and regions surrounding the AR core.
In black we plot the data as taken from \cite{Gary:2001}; see Figure~3 therein.
With a dash-dotted line we show the logarithmic mean between the minimum and maximum values of \cite{Gary:2001} for clarity.
See also \sect{beta}.
\label{F:beta}}
\end{figure}

\cite{Gary:2001} states that one should be careful with constructing magnetic field extrapolations from the photosphere into the corona because when ``moving from the photosphere upwards, \eqi{\beta} can return to \eqi{\sim 1} at relatively low coronal heights, e.g., \eqi{R \sim 1.2}\unit{R_S},'' which are about 140\unit{Mm} in height above the solar surface.
We find here that even this estimate was too conservative and one should rather expect to find regions with \eqi{\beta \sim 1} already at 50\unit{Mm}, which are about 1.07\unit{R_S}, when considering a diverse solar atmosphere, in particular for regions with large variation in magnetic field strength.
One could argue that such upper limit estimates of \eqi{\beta} represent only extreme scenarios, but we find this picture is consistently similar also for the mean values of plasma beta.
We use the logarithmic mean between the minimum and maximum values of the plasma beta in \cite{Gary:2001} (dash-dotted in \fig{beta}) for comparison with our MHD mean \eqi{\beta} value (blue solid line).
While the logarithmic averaged \eqi{\beta} in \cite{Gary:2001} reaches unity at 230\unit{Mm}, we find this happens already as low as 100\unit{Mm} in our computational model.
Above 120\unit{Mm}, the MHD simulation's lowest beta value is even larger than the largest value in \cite{Gary:2001}; we see that the leftmost blue dotted line (MHD minimum \eqi{\beta}) crosses the vertical gray dashed line (\eqi{\beta = 1}) at a lower height than the rightmost dash-dotted line \citep[maximum \eqi{\beta} from][]{Gary:2001} in \fig{beta}.
This result underlines twice the statement from \cite{Gary:2001} that asks for rethinking the paradigm of assuming \eqi{\beta} would be significantly lower than unity everywhere in the corona.
While that region spans from 1 to 146\unit{Mm} in \cite{Gary:2001}, we find it is significantly smaller in a diverse solar atmosphere, where \eqi{\beta} is smaller than unity above 5\unit{Mm} and below 45\unit{Mm} for our 3D MHD model results.

Only recently have \cite{Iwai+al:2014} reported that their observation of coronal magnetic field and plasma pressure leads to a plasma beta that remains inconsistent with the minimum/maximum range given in \cite{Gary:2001}; see the red square in \fig{beta}, where the size of the square indicates the uncertainty on the observed plasma beta value.
Their observational results, on the other hand, fits nicely to our active region data obtained from our MHD model:
Even though their beta estimate is about one order of magnitude below the minimum value given in \cite{Gary:2001} (black dotted line), it is well within our average and minimum values obtained from the active region MHD model data (blue solid and dotted lines).

Also it is worth mentioning that in principle, when we look at our surrounding quiet-Sun area, it is at least theoretically possible that individual atmospheric columns would have a plasma beta strictly larger than unity, see our QS maximum value (orange dotted line in \fig{beta}).
Even though, when we check our model data, we do not find such vertical atmospheric columns.
Hence, the QS maximum beta values originate from different locations or different atmospheric columns.

\section{Rethinking a paradigm, again?} \label{S:paradigm}
At this point, we should address the question: how much do we trust a simulation result?
It is a standing paradigm to trust an observation or a theoretical result more than a computer model.
However, we can indeed rethink this paradigm when a simulation is driven by one type of observation, is checked against other observations for its realism, and keeping in mind that properly solving MHD equations numerically can be as good as an analytic deduction.

A second paradigm is that atmospheric stratifications of solar-like stars so far typically assume a radial magnetic field configuration.
As a consequence of changing this paradigm and also allowing for a varying horizontal field component, one implicitly allows for a more extended chromosphere and a smoother (or wider) transition region to the corona -- in some cases featuring a less steep gradient in the temperature and being located around 3.5\unit{Mm} instead of 2--2.5\unit{Mm} above the photosphere, as proposed by \cite{Bourdin:2014_switch-on}.
The latter and lower value range for the height of the transition region is tied to this paradigm and is supported by multiple works, like the famous VAL-C atmospheric stratification \citep{Vernazza+al:1981} and the FAL-C model \citep{Fontenla+al:1993}.
These previous studies usually rely more on observational or theoretical results -- and none of them is building the bridge between observational and theoretical constraints by the means of a data-driven numerical experiment that has been checked for its realism against {\em distinct} observations, as we like to define a ``trustworthy'' simulation result.

Finally, changing such paradigms influences the energetic requirements to sufficiently heat the corona because inclined fields transport the energy less efficiently in the vertical direction.
This even sets new degrees of freedom for spectro-polarimetric inversions through the use of different atmospheric stratifications, which may yield different magnetic field strengths and hence magnetic vector orientations.
The consequence for seismology in the solar atmosphere is that cutoff frequencies for waves being transmitted through or reflected by various atmospheric layers may deviate from earlier estimates because of the higher variability in plasma beta.

\section{Penumbra formation} \label{S:penumbra}
Within sunspots, the magnetic field is rather vertical and the flux density is about one order of magnitude higher than in quiet-Sun regions.
Therefore, we expect large differences in plasma beta within and outside of sunspots.

\cite{Jurcak:2011} and \cite{Jurcak+al:2015} find that a stable penumbra forms and expands around the umbra of a sunspot, until the umbra--penumbra boundary reaches a typical vertical magnetic field threshold of about \eqi{1860 \pm 190\unit{G}}.
The penumbra is stable if the vertical component remains below the threshold, while the vertical field within the umbra exceeds that threshold value.
Recent MHD models of sunspots, even though they contain only some chromosphere and no corona, already indicate that horizontal magnetic fields, imposed at the upper simulation boundary, lead to a more extended penumbra \citep{Rempel:2012}.
Both findings are not contradictory, instead they suggest that the magnetic field changes from a vertical to a horizontal configuration while the penumbra forms.

In the observation of \cite{Schlichenmaier+al:2012}, we see Doppler red-shifts exist at the edges of forming penumbra patches, exactly where the field inclination changes from horizontal within existing penumbra to a nearly vertical field around it; see \eqi{x=4''} and \eqi{y=7''} between 8:50 and 10:13 in their Figure~2.
This signature of growing penumbra shows that the field inclination becomes more horizontal and simultaneously plasma falls from above.
Yet, it is unclear at this point if the field drags the plasma down or vice versa.
\cite{Schlichenmaier+al:2012} also state that the formation of a stable penumbra avoids regions of emerging magnetic bipoles.
Such avoided regions typically feature upflows; see blue-shifts around \eqi{x=24''} and \eqi{y=7''} in their Figure~2.

\begin{figure}
\plotone{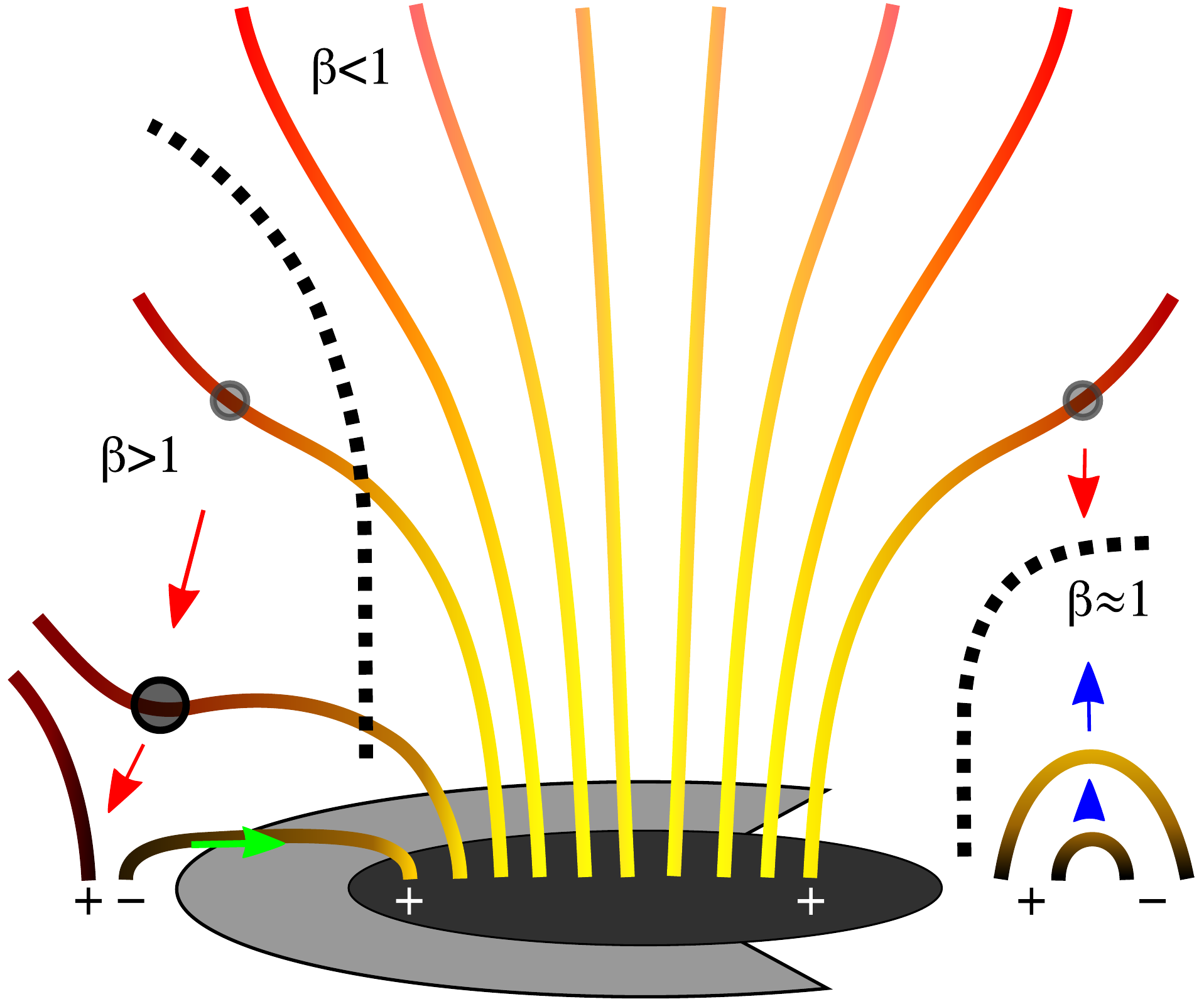}
\caption{Sketch of a sunspot (black) with forming penumbra (gray).
The black dashed line indicates the \eqi{\beta = 1} level, where magnetic and thermal pressure are equal.
Magnetic field lines that are rooted in strong magnetic flux from the sunspot remain vertical within the \eqi{\beta < 1} regime.
Due to the higher magnetic pressure these field lines also remain underdense (yellow).
Slightly inclined field lines on the outer boundaries of the sunspot get loaded with mass through cooling material from above (gray circle).
Gravitation tears down overdense field lines (dark red) that lie in the \eqi{\beta > 1} regime, following the red arrows.
They finally reach the photosphere, magnetically reconnect, and trigger a counter-Evershed flow (green arrow).
Right of the sunspot, emerging flux leads to a plasma beta near unity and hinders penumbra formation because the uprising magnetic structures (blue arrows) give additional support for the overlying inclined field.
See also \sect{penumbra}.
\label{F:sketch}}
\end{figure}

\cite{Shimizu+al:2012} report on observational signatures of an almost horizontal magnetic field in the chromosphere, surrounding the sunspot umbra, where later a penumbra forms in the photosphere.

From our simulation results, summarized in \fig{beta}, we see that \eqi{\beta} is larger than unity in the photosphere for magnetically quiet regions, while it can be expected to be lower than unity for strong flux densities within sunspots, which is similar to the photospheric value range from \cite{Gary:2001}.

We assume now that the variability of \eqi{\beta} is higher than previously estimated in the lower atmosphere.
We consider further that the magnetic connectivity to the corona is sustainably established only if a certain vertical magnetic field strength is reached.
Typically, the magnetic topology of a field extrapolation from a sunspot looks like a funnel-shaped flux tube that expands with height in the chromosphere and transition region. 
This implies that the field becomes more inclined above the surrounding of a sunspot than above its center; see \fig{sketch}.

If some mass from above flows down along the field in a regime where \eqi{\beta > 1}, like coronal material that cools due to radiative losses, this material would eventually reach the \eqi{\beta = 1} layer.
When we also consider that the vertical field experiences less stress through gravity, then the inclined field in the same \eqi{\beta} regime is gravitationally less stable than the more vertical field.

Formulated as a vague idea that is consistent with the observational findings and theoretical considerations mentioned above, inclined field lines may be gravitationally torn down by continuous mass-loading from above -- and this process can start earlier at the boundaries of a sunspot, not only because of the less vertical field inclination but also because of the lower total flux density at the surrounding of a sunspot, where \eqi{\beta} reaches unity higher above the photosphere than directly above the sunspot where the field is stronger and rather vertical; see \fig{sketch}.

Near \eqi{\beta = 1}, magnetic and thermodynamic processes may be equally dominant, which means that when a field line is advected by the plasma, the magnetic connectivity to the surrounding is still significant and such dynamics have a strong influence on neighboring field lines.
Therefore, once a penumbra forms by the magnetic field being dragged down to the photosphere, neighboring field lines also experience an additional force dragging them down.
As a result, one would expect that additional penumbra forms around existing penumbra, which is what is observed.

The bipolar field (see the lower left of \fig{sketch}) originates from a magnetic reconnection process, where field lines that are dragged down to the photosphere encounter small-scale fields in the lower atmospheric layers.
In a consistent manner, \cite{Shimizu+al:2012} report on an observed vertical magnetic field in the photosphere that remains remarkably stable at the outer boundary of the later formed penumbra, as well as the neighboring opposite-polarity field toward the sunspot.
They draw a magnetic canopy structure (see their Figure~5) and observe patches with the same polarity as the sunspot at the location where the footpoints of the canopy are supposed to be.
This proposed canopy structure basically is the result of the falling and reconnecting field-line scenario described along our \fig{sketch}.

Also, \cite{Romano+al:2013} observe several sites of opposite-polarity patches associated with chromospheric downflows around a pore.
The authors conclude additional support for a canopy-like magnetic structure in the chromosphere before the penumbra forms in the photosphere.
Consistent with this study, \cite{Romano+al:2014} associate the penumbra formation with an observed change in the magnetic field inclination around a pore.

The mechanism for penumbra formation proposed here relies on initially inclined magnetic field.
Even though one can always expect such field around a strong flux concentration, like a sunspot, the penumbra formation through this mechanism can be expected to set in preferably where the field is initially most inclined.
\cite{Rezaei+al:2012} suggest that a reconfiguration mechanism may lead to more inclined magnetic field at the edge of a sunspot light bridge.
As expected, in their observation the penumbra starts forming exactly there.

The counter-Evershed flow, that is inwards to the sunspot, observed during penumbra formation \citep{Schlichenmaier+al:2011} can be understood as a result of a reconnection process between a falling field line and pre-existing magnetic field in the lower atmosphere because these field lines are initially connected to lower-pressure atmospheric layers.
These field lines need to be filled with additional mass when reconnecting from a lower thermal-pressure regime (in the upper atmosphere) to photospheric layers outside of the sunspot.
The equilibration of this pressure imbalance requires a short-lived plasma flow along the penumbral field toward the sunspot during the early penumbra formation phase.
Reconnection outflows along the separatrix field lines \citep[see][]{Bourdin:2017} may additionally drive the observed counter-Evershed flows; see the green arrow in \fig{sketch}.

\cite{Romano+al:2014} find small-scale magnetic patches of opposite polarity with an upflow farther outside and a downflow next to the pore.
This observational signature can be expected from a flux tube that just has reconnected to the photosphere and then equilibrates a pressure difference between near and away from the pore by a field-aligned plasma flow, see green arrow in \fig{sketch}.

To further check this idea one should find a correlation of chromospheric redshifts on the order of some \unit{km/s} above the edges and shortly before a forming penumbra, similar to that shown in \cite{Schlichenmaier+al:2012} -- eventually followed by a counter-Evershed flow that originates in a newly created opposite-polarity patch, similar to the observation in \cite{Romano+al:2014}.

\section{Conclusions} \label{S:conclusions}

Standard solar atmospheric models, as established by \cite{Vernazza+al:1981} and \cite{Fontenla+al:1993}, are derived from estimates of the photospheric magnetic field.
The famous VAL-C model is in this case applicable for a rather strong surface flux density of 2000\unit{Gauss} or more, which represents the umbra of a sunspot.
There, the further assumption of a rather vertical magnetic field is valid, but this represents only about 2\unit{\%} of the solar surface.
If we consider the large variation in magnetic field inclinations and surface flux densities, which then represents the whole solar surface, we find significantly different atmospheric columns than proposed and used in most earlier works.
This ultimately leads to significantly less volume of \eqi{\beta} being lower than unity (and hence with magnetically confined plasma) in the corona than previously thought \citep[see][]{Gary:2001}.

As a result, a penumbra may form through field lines that are filled by cooling material downflowing from the corona and eventually dragging the field down to the photosphere by gravitation.
This process requires an initially strong vertical field in order to establish a sufficiently long existing connectivity to the corona.
Hence, the \cite{Jurcak:2011} criterion of a typical vertical field threshold at the umbra--penumbra boundary can be explained by such a process that becomes possible when considering \eqi{\beta} values larger and smaller than unity within the lower atmosphere, as our 3D~MHD simulation results suggest.
During this process we expect counter-Evershed flows in field lines that magnetically reconnect from the upper atmosphere to the photosphere.

A Doppler-shift pattern that matches the proposed pressure equilibration after magnetic reconnection from higher to lower atmospheric layers also is observed near a pore when the penumbra formation sets in \citep{Murabito+al:2016}.

We also find that gradients in the temperature stratification are significantly less strong in the quiet Sun than directly above strong magnetic polarities with rather vertical magnetic fields.
This reduces the energetic coronal heating requirements above these quiet areas, as compared to active regions.
When considering a mixture of weaker and stronger magnetic fields in the photosphere, this yields a wider and less steep transition region for stellar atmospheric stratifications in general.


\acknowledgments

PB thanks the referee for helpful hints.
The results of this research have been achieved using the PRACE Research Infrastructure resource \emph{Curie} based in France at TGCC, as well as \emph{JuRoPA} hosted by the J{\"u}lich Supercomputing Centre in Germany.
{\em Hinode} is a Japanese mission developed, launched, and operated by ISAS/JAXA, in partnership with NAOJ, NASA, and STFC (UK). Additional operational support is provided by ESA and NSC (Norway).
The {\em STEREO}/SECCHI data used here were produced by an international consortium of the Naval Research Laboratory (USA), Lockheed Martin Solar and Astrophysics Lab (USA), NASA Goddard Space Flight Center (USA), Rutherford Appleton Laboratory (UK), University of Birmingham (UK), Max-Planck-Institut f\"ur Sonnensystemforschung (Germany), Centre Spatiale de Li\`ege (Belgium), Institut d'Optique Th\'eorique et Appliqu\'ee (France), and Institut d'Astrophysique Spatiale (France).

\bibliography{Literatur}

\begin{thebibliography}{27}
\providecommand{\natexlab}[1]{#1}
\providecommand{\url}[1]{\texttt{#1}}
\expandafter\ifx\csname urlstyle\endcsname\relax
  \providecommand{\doi}[1]{doi: #1}\else
  \providecommand{\doi}{doi: \begingroup \urlstyle{rm}\Url}\fi

\bibitem[{Bello Gonz{\'a}lez} et~al.(in print){Bello Gonz{\'a}lez}, {Jur{\v
  c}{\'a}k}, {Schlichenmaier}, and {Rezaei}]{Bello_Gonzales+al:2017}
N.~{Bello Gonz{\'a}lez}, J.~{Jur{\v c}{\'a}k}, R.~{Schlichenmaier}, and
  R.~{Rezaei}.
\newblock {New insights on penumbra magneto-convection}.
\newblock In Luca {Belluzzi}, editor, \emph{SPW8 Florence Sept. 2016},
  Astronomical Society of the Pacific Conference Series, in print.

\bibitem[Bourdin(2014)]{Bourdin:2014_switch-on}
Ph.-A. Bourdin.
\newblock {Standard 1D solar atmosphere as initial condition for MHD
  simulations and switch-on effects}.
\newblock \emph{\ceab}, 38\penalty0 (1):\penalty0 1--10, dec 2014.

\bibitem[{Bourdin}(2017)]{Bourdin:2017}
Ph.-A. {Bourdin}.
\newblock {Catalog of fine-structured electron velocity distribution functions
  -- Part 1: Antiparallel magnetic-field reconnection (Geospace Environmental
  Modeling case)}.
\newblock \emph{\angeo}, 35:\penalty0 1051--1067, sep 2017.
\newblock \doi{10.5194/angeo-35-1-2017}.

\bibitem[Bourdin et~al.(2013)Bourdin, Bingert, and
  Peter]{Bourdin+al:2013_overview}
Ph.-A. Bourdin, S.~Bingert, and H.~Peter.
\newblock {Observationally driven 3D~MHD model of the solar corona above an
  active region}.
\newblock \emph{\aap}, 555\penalty0 (A123):\penalty0 A123 (6pp), July 2013.
\newblock \doi{10.1051/0004-6361/201321185}.

\bibitem[Bourdin et~al.(2014)Bourdin, Bingert, and
  Peter]{Bourdin+al:2014_coronal-loops}
Ph.-A. Bourdin, S.~Bingert, and H.~Peter.
\newblock {Coronal loops above an Active Region: Observation versus model}.
\newblock \emph{\pasj}, 66\penalty0 (S7):\penalty0 1--8, dec 2014.
\newblock \doi{10.1093/pasj/psu123}.

\bibitem[Bourdin et~al.(2015)Bourdin, Bingert, and
  Peter]{Bourdin+al:2015_energy-input}
Ph.-A. Bourdin, S.~Bingert, and H.~Peter.
\newblock {Coronal energy input and dissipation in a solar Active Region 3D~MHD
  model}.
\newblock \emph{\aap}, 580\penalty0 (A72):\penalty0 A72 (8pp), aug 2015.
\newblock \doi{10.1051/0004-6361/201525839}.

\bibitem[Bourdin et~al.(2016)Bourdin, Bingert, and
  Peter]{Bourdin+al:2016_scaling-laws}
Ph.-A. Bourdin, S.~Bingert, and H.~Peter.
\newblock {Scaling laws of coronal loops compared to a 3D~MHD model of an
  active region}.
\newblock \emph{\aap}, 589:\penalty0 A86, may 2016.
\newblock \doi{10.1051/0004-6361/201525840}.

\bibitem[{Cook} et~al.(1989){Cook}, {Cheng}, {Jacobs}, and
  {Antiochos}]{Cook+al:1989}
J.~W. {Cook}, C.-C. {Cheng}, V.~L. {Jacobs}, and S.~K. {Antiochos}.
\newblock {Effect of coronal elemental abundances on the radiative loss
  function}.
\newblock \emph{\apj}, 338:\penalty0 1176--1183, March 1989.
\newblock \doi{10.1086/167268}.

\bibitem[{Culhane} et~al.(2007){Culhane}, {Harra}, {James}, {Al-Janabi},
  {Bradley}, {Chaudry}, {Rees}, {Tandy}, {Thomas}, {Whillock}, {Winter},
  {Doschek}, {Korendyke}, {Brown}, {Myers}, {Mariska}, {Seely}, {Lang}, {Kent},
  {Shaughnessy}, {Young}, {Simnett}, {Castelli}, {Mahmoud}, {Mapson-Menard},
  {Probyn}, {Thomas}, {Davila}, {Dere}, {Windt}, {Shea}, {Hagood}, {Moye},
  {Hara}, {Watanabe}, {Matsuzaki}, {Kosugi}, {Hansteen}, and
  {Wikstol}]{Culhane+al:2007}
J.~L. {Culhane}, L.~K. {Harra}, A.~M. {James}, K.~{Al-Janabi}, L.~J. {Bradley},
  R.~A. {Chaudry}, K.~{Rees}, J.~A. {Tandy}, P.~{Thomas}, M.~C.~R. {Whillock},
  B.~{Winter}, G.~A. {Doschek}, C.~M. {Korendyke}, C.~M. {Brown}, S.~{Myers},
  J.~{Mariska}, J.~{Seely}, J.~{Lang}, B.~J. {Kent}, B.~M. {Shaughnessy}, P.~R.
  {Young}, G.~M. {Simnett}, C.~M. {Castelli}, S.~{Mahmoud}, H.~{Mapson-Menard},
  B.~J. {Probyn}, R.~J. {Thomas}, J.~{Davila}, K.~{Dere}, D.~{Windt},
  J.~{Shea}, R.~{Hagood}, R.~{Moye}, H.~{Hara}, T.~{Watanabe}, K.~{Matsuzaki},
  T.~{Kosugi}, V.~{Hansteen}, and {\O}.~{Wikstol}.
\newblock {The EUV Imaging Spectrometer for Hinode}.
\newblock \emph{\solphys}, 243:\penalty0 19--61, June 2007.
\newblock \doi{10.1007/s01007-007-0293-1}.

\bibitem[Fontenla et~al.(1993)Fontenla, Avrett, and Loeser]{Fontenla+al:1993}
J.~M. Fontenla, E.~H. Avrett, and R.~Loeser.
\newblock {Energy Balance in the Solar Transition Region. III. Helium Emission
  in Hydrostatic, Constant Abundance Models with Diffusion}.
\newblock \emph{\apj}, 406:\penalty0 319--345, 1993.

\bibitem[{Gary}(2001)]{Gary:2001}
G.~A. {Gary}.
\newblock {Plasma Beta above a Solar Active Region: Rethinking the Paradigm}.
\newblock \emph{\solphys}, 203:\penalty0 71--86, October 2001.
\newblock \doi{10.1023/A:1012722021820}.

\bibitem[{Howard} et~al.(2008){Howard}, {Moses}, {Vourlidas}, {Newmark},
  {Socker}, {Plunkett}, {Korendyke}, {Cook}, {Hurley}, {Davila}, {Thompson},
  {St Cyr}, {Mentzell}, {Mehalick}, {Lemen}, {Wuelser}, {Duncan}, {Tarbell},
  {Wolfson}, {Moore}, {Harrison}, {Waltham}, {Lang}, {Davis}, {Eyles},
  {Mapson-Menard}, {Simnett}, {Halain}, {Defise}, {Mazy}, {Rochus}, {Mercier},
  {Ravet}, {Delmotte}, {Auchere}, {Delaboudiniere}, {Bothmer}, {Deutsch},
  {Wang}, {Rich}, {Cooper}, {Stephens}, {Maahs}, {Baugh}, {McMullin}, and
  {Carter}]{Howard+al:2008}
R.~A. {Howard}, J.~D. {Moses}, A.~{Vourlidas}, J.~S. {Newmark}, D.~G. {Socker},
  S.~P. {Plunkett}, C.~M. {Korendyke}, J.~W. {Cook}, A.~{Hurley}, J.~M.
  {Davila}, W.~T. {Thompson}, O.~C. {St Cyr}, E.~{Mentzell}, K.~{Mehalick},
  J.~R. {Lemen}, J.~P. {Wuelser}, D.~W. {Duncan}, T.~D. {Tarbell}, C.~J.
  {Wolfson}, A.~{Moore}, R.~A. {Harrison}, N.~R. {Waltham}, J.~{Lang}, C.~J.
  {Davis}, C.~J. {Eyles}, H.~{Mapson-Menard}, G.~M. {Simnett}, J.~P. {Halain},
  J.~M. {Defise}, E.~{Mazy}, P.~{Rochus}, R.~{Mercier}, M.~F. {Ravet},
  F.~{Delmotte}, F.~{Auchere}, J.~P. {Delaboudiniere}, V.~{Bothmer},
  W.~{Deutsch}, D.~{Wang}, N.~{Rich}, S.~{Cooper}, V.~{Stephens}, G.~{Maahs},
  R.~{Baugh}, D.~{McMullin}, and T.~{Carter}.
\newblock {Sun Earth Connection Coronal and Heliospheric Investigation
  (SECCHI)}.
\newblock \emph{\ssr}, 136:\penalty0 67--115, April 2008.
\newblock \doi{10.1007/s11214-008-9341-4}.

\bibitem[{Iwai} et~al.(2014){Iwai}, {Shibasaki}, {Nozawa}, {Takahashi},
  {Sawada}, {Kitagawa}, {Miyawaki}, and {Kashiwagi}]{Iwai+al:2014}
K.~{Iwai}, K.~{Shibasaki}, S.~{Nozawa}, T.~{Takahashi}, S.~{Sawada},
  J.~{Kitagawa}, S.~{Miyawaki}, and H.~{Kashiwagi}.
\newblock {Coronal magnetic field and the plasma beta determined from radio and
  multiple satellite observations}.
\newblock \emph{\eps}, 66:\penalty0 149, December 2014.
\newblock \doi{10.1186/s40623-014-0149-z}.

\bibitem[{Jur{\v c}{\'a}k}(2011)]{Jurcak:2011}
J.~{Jur{\v c}{\'a}k}.
\newblock {Azimuthal variations of magnetic field strength and inclination on
  penumbral boundaries}.
\newblock \emph{\aap}, 531:\penalty0 A118, July 2011.
\newblock \doi{10.1051/0004-6361/201015959}.

\bibitem[{Jur{\v c}{\'a}k} et~al.(2015){Jur{\v c}{\'a}k}, {Bello Gonz{\'a}lez},
  {Schlichenmaier}, and {Rezaei}]{Jurcak+al:2015}
J.~{Jur{\v c}{\'a}k}, N.~{Bello Gonz{\'a}lez}, R.~{Schlichenmaier}, and
  R.~{Rezaei}.
\newblock {A distinct magnetic property of the inner penumbral boundary.
  Formation of a stable umbra-penumbra boundary in a sunspot}.
\newblock \emph{\aap}, 580:\penalty0 L1, August 2015.
\newblock \doi{10.1051/0004-6361/201425501}.

\bibitem[{Kosugi} et~al.(2007){Kosugi}, {Matsuzaki}, {Sakao}, {Shimizu},
  {Sone}, {Tachikawa}, {Hashimoto}, {Minesugi}, {Ohnishi}, {Yamada}, {Tsuneta},
  {Hara}, {Ichimoto}, {Suematsu}, {Shimojo}, {Watanabe}, {Shimada}, {Davis},
  {Hill}, {Owens}, {Title}, {Culhane}, {Harra}, {Doschek}, and
  {Golub}]{Kosugi+al:2007}
T.~{Kosugi}, K.~{Matsuzaki}, T.~{Sakao}, T.~{Shimizu}, Y.~{Sone},
  S.~{Tachikawa}, T.~{Hashimoto}, K.~{Minesugi}, A.~{Ohnishi}, T.~{Yamada},
  S.~{Tsuneta}, H.~{Hara}, K.~{Ichimoto}, Y.~{Suematsu}, M.~{Shimojo},
  T.~{Watanabe}, S.~{Shimada}, J.~M. {Davis}, L.~D. {Hill}, J.~K. {Owens},
  A.~M. {Title}, J.~L. {Culhane}, L.~K. {Harra}, G.~A. {Doschek}, and
  L.~{Golub}.
\newblock {The Hinode (Solar-B) Mission: An Overview}.
\newblock \emph{\solphys}, 243:\penalty0 3--17, June 2007.
\newblock \doi{10.1007/s11207-007-9014-6}.

\bibitem[{Murabito} et~al.(2016){Murabito}, {Romano}, {Guglielmino},
  {Zuccarello}, and {Solanki}]{Murabito+al:2016}
M.~{Murabito}, P.~{Romano}, S.~L. {Guglielmino}, F.~{Zuccarello}, and S.~K.
  {Solanki}.
\newblock {Formation of the Penumbra and Start of the Evershed Flow}.
\newblock \emph{\apj}, 825:\penalty0 75, July 2016.
\newblock \doi{10.3847/0004-637X/825/1/75}.

\bibitem[{Rempel}(2012)]{Rempel:2012}
M.~{Rempel}.
\newblock {Numerical Sunspot Models: Robustness of Photospheric Velocity and
  Magnetic Field Structure}.
\newblock \emph{\apj}, 750:\penalty0 62, May 2012.
\newblock \doi{10.1088/0004-637X/750/1/62}.

\bibitem[{Rezaei} et~al.(2012){Rezaei}, {Bello Gonz{\'a}lez}, and
  {Schlichenmaier}]{Rezaei+al:2012}
R.~{Rezaei}, N.~{Bello Gonz{\'a}lez}, and R.~{Schlichenmaier}.
\newblock {The formation of sunspot penumbra. Magnetic field properties}.
\newblock \emph{\aap}, 537:\penalty0 A19, January 2012.
\newblock \doi{10.1051/0004-6361/201117485}.

\bibitem[{Romano} et~al.(2013){Romano}, {Frasca}, {Guglielmino}, {Ermolli},
  {Tritschler}, {Reardon}, and {Zuccarello}]{Romano+al:2013}
P.~{Romano}, D.~{Frasca}, S.~L. {Guglielmino}, I.~{Ermolli}, A.~{Tritschler},
  K.~P. {Reardon}, and F.~{Zuccarello}.
\newblock {Velocity and Magnetic Field Distribution in a Forming Penumbra}.
\newblock \emph{\apjl}, 771:\penalty0 L3, July 2013.
\newblock \doi{10.1088/2041-8205/771/1/L3}.

\bibitem[{Romano} et~al.(2014){Romano}, {Guglielmino}, {Cristaldi}, {Ermolli},
  {Falco}, and {Zuccarello}]{Romano+al:2014}
P.~{Romano}, S.~L. {Guglielmino}, A.~{Cristaldi}, I.~{Ermolli}, M.~{Falco}, and
  F.~{Zuccarello}.
\newblock {Evolution of the Magnetic Field Inclination in a Forming Penumbra}.
\newblock \emph{\apj}, 784:\penalty0 10, March 2014.
\newblock \doi{10.1088/0004-637X/784/1/10}.

\bibitem[{Schlichenmaier} et~al.(2011){Schlichenmaier}, {Gonz{\'a}lez}, and
  {Rezaei}]{Schlichenmaier+al:2011}
R.~{Schlichenmaier}, N.~B. {Gonz{\'a}lez}, and R.~{Rezaei}.
\newblock {The formation of a penumbra as observed with the German VTT and
  SoHO/MDI}.
\newblock In D.~{Prasad Choudhary} and K.~G. {Strassmeier}, editors,
  \emph{Physics of Sun and Star Spots}, volume~6 of \emph{Proceedings of the
  International Astronomical Union, IAU Symposium 273}, pages 134--140, August
  2011.
\newblock \doi{10.1017/S1743921311015134}.

\bibitem[{Schlichenmaier} et~al.(2012){Schlichenmaier}, {Rezaei}, and
  {Gonz{\'a}lez}]{Schlichenmaier+al:2012}
R.~{Schlichenmaier}, R.~{Rezaei}, and N.~B. {Gonz{\'a}lez}.
\newblock {On the Formation of Penumbrae as Observed with the German VTT
  SOHO/MDI, and SDO/HMI}.
\newblock In L.~{Bellot Rubio}, F.~{Reale}, and M.~{Carlsson}, editors,
  \emph{4th Hinode Science Meeting: Unsolved Problems and Recent Insights},
  volume 455 of \emph{Astronomical Society of the Pacific Conference Series},
  page~61, May 2012.

\bibitem[{Shimizu} et~al.(2012){Shimizu}, {Ichimoto}, and
  {Suematsu}]{Shimizu+al:2012}
T.~{Shimizu}, K.~{Ichimoto}, and Y.~{Suematsu}.
\newblock {Precursor of Sunspot Penumbral Formation Discovered with Hinode
  Solar Optical Telescope Observations}.
\newblock \emph{\apjl}, 747:\penalty0 L18, March 2012.
\newblock \doi{10.1088/2041-8205/747/2/L18}.

\bibitem[Spitzer(1962)]{Spitzer:1962}
L.~Spitzer.
\newblock \emph{{Physics of Fully Ionized Gases}}.
\newblock Interscience, New York (2nd edition), 1962.

\bibitem[{Tsuneta} et~al.(2008){Tsuneta}, {Ichimoto}, {Katsukawa}, {Nagata},
  {Otsubo}, {Shimizu}, {Suematsu}, {Nakagiri}, {Noguchi}, {Tarbell}, {Title},
  {Shine}, {Rosenberg}, {Hoffmann}, {Jurcevich}, {Kushner}, {Levay}, {Lites},
  {Elmore}, {Matsushita}, {Kawaguchi}, {Saito}, {Mikami}, {Hill}, and
  {Owens}]{Tsuneta+al:2008}
S.~{Tsuneta}, K.~{Ichimoto}, Y.~{Katsukawa}, S.~{Nagata}, M.~{Otsubo},
  T.~{Shimizu}, Y.~{Suematsu}, M.~{Nakagiri}, M.~{Noguchi}, T.~{Tarbell},
  A.~{Title}, R.~{Shine}, W.~{Rosenberg}, C.~{Hoffmann}, B.~{Jurcevich},
  G.~{Kushner}, M.~{Levay}, B.~{Lites}, D.~{Elmore}, T.~{Matsushita},
  N.~{Kawaguchi}, H.~{Saito}, I.~{Mikami}, L.~D. {Hill}, and J.~K. {Owens}.
\newblock {The Solar Optical Telescope for the Hinode Mission: An Overview}.
\newblock \emph{\solphys}, 249:\penalty0 167--196, June 2008.
\newblock \doi{10.1007/s11207-008-9174-z}.

\bibitem[Vernazza et~al.(1981)Vernazza, Avrett, and Loeser]{Vernazza+al:1981}
J.~E. Vernazza, E.~H. Avrett, and R.~Loeser.
\newblock {Structure of the solar chromosphere. III. Models of the EUV
  brightness components of the quiet Sun}.
\newblock \emph{\apjs}, 45:\penalty0 635--725, 1981.

\end{thebibliography}
\bibliographystyle{plainnat}

\end{document}